\documentstyle[epsf,12pt]{article}

\renewcommand{\baselinestretch}{1.23}

\begin{document}

\def\be{\begin{eqnarray}}
\def\en{\end{eqnarray}}
\def\non{\nonumber}
\def\la{\langle}
\def\ra{\rangle}
\def\up{\uparrow}
\def\down{\downarrow}
\def\tr{{\rm tr}}
\def\ep{\varepsilon}
\def\lsim{ {\ \lower-1.2pt\vbox{\hbox{\rlap{$<$}\lower5pt\vbox{\hbox{$\sim$}
}}}\ } }
\def\gsim{ {\ \lower-1.2pt\vbox{\hbox{\rlap{$>$}\lower5pt\vbox{\hbox{$\sim$}
}}}\ } }
\def\pr{{\sl Phys. Rev.}~}
\def\prl{{\sl Phys. Rev. Lett.}~}
\def\pl{{\sl Phys. Lett.}~}
\def\np{{\sl Nucl. Phys.}~}
\def\zp{{\sl Z. Phys.}~}

\title{
\begin{flushright}
{\normalsize IP-ASTP-01-97}
\end{flushright}
\vskip 10mm
\Large\bf
Remarks on the Strong Coupling Constants in Heavy Hadron Chiral Lagrangians 
}
\author{{\bf Hai-Yang Cheng}\\
{\it Institute of Physics, Academia Sinica}\\
{\it Taipei, Taiwan 115, Republic of China}\\
}
\date{(January, 1997)}

\maketitle

\begin{abstract}
  In the heavy quark limit,
   there are three independent strong axial coupling constants in chiral
Lagrangians for ground-state heavy hadrons:
$g$ in the meson sector, and $g_1,~g_2$ in the baryon sector.
The coupling $g_2$ is extracted from recent CLEO measurements of 
$\Sigma_c^*\to\Lambda_c\pi$ to be $|g_2|=0.57\pm 0.10$, in agreement with
the nonrelativistic quark model expectation but in $2\sigma$ deviation from
the large-$N_c$ argument. The parameter $g_1$ cannot be determined
directly from strong decays of heavy baryons. Nevertheless, some information
of $g_1$ can be learned from the radiative decay $\Xi_c^{'*0}\to\Xi_c^0
\gamma$, which is prohibited at tree level by SU(3) symmetry but induced 
by chiral loops.
A measurement of $\Gamma(\Xi_c^{'*0}\to\Xi_c^0\gamma)$ will yield two possible
solutions for $g_1$.  Assuming the validity of the quark model relations
among different coupling constants, the experimental value of $g_2$ 
implies that $|g|=0.70\pm 0.12$ and $|g_1|=0.93\pm 0.16$\,.

\end{abstract}
\pagebreak

    A most suitable framework for studying the low-energy dynamics of heavy
hadrons is provided by the formalism in which heavy quark symmetry and
chiral symmetry are synthesized \cite{Yan,Wise,Dono}. In the limit of heavy
quark symmetry, the heavy hadron chiral Lagrangian for strong interactions
consists of three unknown strong coupling constants in the low-energy
interactions between the Goldstone bosons and ground-state heavy hadrons: $g$ 
in the meson sector, and $g_1,~g_2$ in the baryon sector. These three coupling
constants are independent of the heavy quark species involved. In principle,
the decays $D^*\to D\pi,~\Sigma_c^*\to\Sigma_c\pi$ and $\Sigma_c\to\Lambda_c
\pi$ can be utilized to determine the parameters $g,~g_1$ and $g_2$, 
respectively.
Unfortunately, none of the decay rates of above decays has been measured.
In particular, the decay $\Sigma_c^*\to\Sigma_c\pi$ (or $\Omega_c^*\to
\Omega_c\pi$) is nowadays known to be kinematically prohibited since the mass 
difference between $\Sigma_c^*$ and $\Sigma_c$ is only of order 65 MeV. 
This means that the coupling constant $g_1$ cannot be determined directly 
from the strong decays of heavy
baryons. Nevertheless, as pointed out in \cite{Yan}, the parameters $g_1$ 
and $g_2$ are related to $g$ in the nonrelativistic quark model via
\footnote{The relation $g_1={1\over 3}g$ originally given in \cite{Yan} is
erroneous; the correction will appear in an erratum.}
\be
g_1=\,{4\over 3}g,~~~~~~g_2=-\sqrt{2\over 3}\,g.
\en
In this short Letter we will extract $g_2$ from the recent CLEO measurement 
of $\Sigma_c^*\to\Lambda_c\pi$ decays.
We then discuss its implications and the feasibility of measuring $g_1$ from
the radiative decay $\Xi_c^{'*0}\to\Xi_c^0\gamma$.

   The most general chiral-invariant low-energy interactions of ground-state
heavy hadrons with the Goldstone bosons are given by (we follow the notation
of \cite{Yan})
\be
{\cal L}_{\rm int} &=& f\sqrt{m_Pm_{P^*}}\,(PA^\mu P_\mu^{*\dagger}+P_\mu^*
A^\mu P^\dagger)+{1\over 2}g\epsilon_{\mu\nu\lambda\kappa}(P^{*\mu\nu}
A^\lambda P^{*\kappa\dagger}+P^{*\kappa}A^\lambda P^{*\mu\nu\dagger})  \non\\
&+& g_1\tr(\bar B_6\gamma_\mu\gamma_5 A^\mu
B_6)+g_2\tr(\bar B_6\gamma_\mu\gamma_5A^\mu B_{\bar 3})+~{\rm h.c.}   \\
&+& g_3\tr(\bar B^*_{6\mu}A^\mu B_6)+~{\rm h.c.}+g_4\tr(\bar B^{*\mu}_6A_\mu
B_{\bar 3})+~{\rm h.c.}   \non \\
&+& g_5\tr(\bar B_6^{*\nu}\gamma_\mu\gamma_5A^\mu B^*_{6\nu})+g_6\tr(\bar
B_{\bar 3}\gamma_\mu\gamma_5 A^\mu B_{\bar 3}),   \non
\en
with $P^{*\dagger}_{\mu\nu}=D_\mu P_\nu^{*\dagger}-D_\nu P_\mu^{*\dagger}$,
where $P$ and $P^*$ denote the ground-state $0^-$ and $1^-$ heavy mesons,
respectively, $B_{\bar 3}$ the spin-${1\over 2}$ antitriplet baryon, $B_6$
the spin-${1\over 2}$ sextet baryon, $B_6^*$ the spin-${3\over 2}$ baryon,
and $V_\mu,~A_\mu$ are, respectively, the chiral vector and 
axial-vector fields. In the infinite heavy quark mass limit, heavy 
quark symmetry reduces the eight coupling constants to three independent
ones, say $g,~g_1$ and $g_2$ \cite{Yan}:\footnote{ In \cite{Cho} the three 
independent parameters are denoted by $g_1,~g_2,~g_3$, which are related 
to our notation by
\be
g_1^{\rm Cho}=g,~~~g_2^{\rm Cho}=-{3\over 2}g_1,~~~g_3^{\rm Cho}=-\sqrt{3}
g_2.   \non
\en}
\be
f=2g,~~~g_3={\sqrt{3}\over 2}g_1,~~~g_5=-{3\over 2}g_1,~~~g_4=-\sqrt{3}g_2,
~~~g_6=0.
\en
Furthermore, there is only one independent coupling constant within the 
context of the nonrelativistic quark model [see Eq.~(1)]. 
\footnote{Using spin-flavor relativistic supermultiplet theory, it is shown
in \cite{Delb} that all the strong and electromagnetic interactions of
ground-state hadrons can be expressed as a single coupling constant. However,
the predicted strong decay widths of charmed baryons are in general quite 
different from 
ours. For example, the calculated rate $\Gamma(\Sigma_c\to\Lambda_c\pi)=28$
keV in \cite{Delb} is two orders of magnitude smaller than ours [see
Eq.~(17) below].}
In the presence of $1/m_Q$ corrections, the heavy quark symmetry 
relations (3) are violated by the $1/m_Q$ heavy quark chromomagnetic operator
$O_2\sim \bar Q\sigma\cdot GQ$. Nevertheless, the $O_2$-induced $1/m_Q$ 
corrections to the coupling constants still satisfy certain model-independent
relations as elaborated in \cite{1/M}.

   The theoretical estimate of the meson strong coupling constant $g$ is
rather diverse. Experimentally, the only constraint comes from the upper
limit $\Gamma(D^{*+})<131$ keV set by ACCMOR Collaboration \cite{ACCMOR},
which leads to $g<0.74$\,. Theoretically, it is related in the nonrelativistic
quark model to $g_A^{ud}$, the axial-vector coupling in the single quark 
transition $u\to d$, which has the value of 0.75 if the same model is 
required to produce the correct value of $g_A^N\simeq 1.25$ \cite{Yan}. 
Relativistic effects of quark motion and quark spins are expected to reduce
the value of $g$. Using the relativistic light-front quark model \cite{CCH}
we find $g\sim 0.50-0.55$\,, which is consistent with \cite{Donnel}. It has 
also been estimated in the approach of the QCD sum rule with the result 
\cite{QSR}: $g\sim 0.21-0.39\,$. Using $g_A^{ud}=0.75$, the coupling
constants
\be
g=0.75\,,~~~~g_1=1\,,~~~~g_2=-0.61
\en
can be regarded as the benchmarked values of $g_i$ in the nonrelativistic
quark model \cite{Yan}. It should be stressed that, in spite of the
experimental constraint $g<0.74$\,, quark model calculations of the 
branching ratios of strong and radiative decays of charmed mesons based on
$g=0.75$ are consistent with experiment \cite{em}. In the large-$N_c$ 
approach, the baryon coupling constants are related to the nucleon axial
coupling $g_A^N$ via \cite{Manohar}
\be
g_1=\,g_A^N,~~~~~g_2=-{1\over\sqrt{2}}\,g_A^N.
\en

  The recent CLEO measurement of $\Sigma_c^*\to\Lambda_c\pi^\pm$ decays allows
us to determine the coupling $g_2$ directly. From the baryon matrices
\be
B_6=\left(\matrix{ \Sigma_c^{++} & {1\over\sqrt{2}}\Sigma_c^+ & 
{1\over\sqrt{2}}\Xi'^+_c \cr   {1\over\sqrt{2}}\Sigma_c^+ & \Sigma_c^0  &
{1\over\sqrt{2}}\Xi'^0_c   \cr   {1\over\sqrt{2}}\Xi'^+_c & {1\over\sqrt{2}}
\Xi'^0_c & \Omega_c^0  \cr}\right),~~~~~~B_{\bar 3}=\left(\matrix{ 0 & 
\Lambda^+_c & \Xi_c^+  \cr -\Lambda_c^+ & 0 & \Xi_c^0   \cr  -\Xi_c^+ & 
-\Xi_c^0   & 0 \cr}\right),
\en
and $B_{6}^*$ similar to $B_6$, and
\be
A_\mu=-{1\over f_\pi}\,\partial_\mu({1\over 2}\tau^a\pi^a)+\cdots,
\en
we obtain
\be
A(\Sigma_c^*\to\Lambda_c\pi)=-i{\sqrt{3}\over \sqrt{2}}\,{g_2\over f_\pi}
\,\bar u_{\Lambda_c}q_\mu u^\mu_{\Sigma_c^*},~~~
A(\Sigma_c\to\Lambda_c\pi)=-i{\sqrt{3}\over \sqrt{2}}\,{g_2\over f_\pi}
\,\bar u_{\Lambda_c}q_\mu u_{\Sigma_c},
\en
where $u^\mu$ is a Rarita-Schwinger vector-spinor for a spin-${3\over 2}$ 
particle, $q_\mu$ is the pion 4-momentum,
and $f_\pi=93$ MeV. It is straightforward to show that
\be
\Gamma(\Sigma_c^*(v)\to\Lambda_c(v)\pi)=\,{p_c^3\over 6\pi}\left({\sqrt{3}g_2
\over \sqrt{2}f_\pi}\right)^2\,{m_{\Lambda_c}\over m_{\Sigma_c^*}}, \non\\
\Gamma(\Sigma_c(v)\to\Lambda_c(v)\pi)=\,{p_c^3\over 6\pi}\left({\sqrt{3}g_2
\over \sqrt{2}f_\pi}\right)^2\,{m_{\Lambda_c}\over m_{\Sigma_c}},
\en
where $p_c$ is the c.m. momentum of the final-state particle. 
In heavy quark effective theory, the velocity of $\Lambda_c$ is taken to be 
the same as that of $\Sigma_c^*$ or $\Sigma_c$. However, we will not apply
heavy quark symmetry to the phase space due to large $1/m_Q$ corrections to
$p_c^3$ ($m_{\Sigma_c}=m_{\Sigma_c^*}$ in heavy quark limit).
If the realistic value of $v_{\Lambda_c}$ is employed,
the decay rates will become
\be
\Gamma(\Sigma_c^*(v)\to\Lambda_c(v')\pi)&=&{p_c^3\over 6\pi}\left({\sqrt{3}g_2
\over \sqrt{2}f_\pi}\right)^2\,{(m_{\Sigma_c^*}+m_{\Lambda_c})^2-m_\pi^2\over 
4m^2_{\Sigma_c^*}},   \non \\
\Gamma(\Sigma_c(v)\to \Lambda_c(v')\pi) &=& {p^3_c\over 6\pi}\left({\sqrt{3}
g_2\over \sqrt{2}f_\pi}\right)^2.
\en
It is easy to check that the zero-recoil relation $v\cdot v'=1$ is a 
very good approximation for $\Sigma_c^*\to\Lambda_c\pi$ decay. As a 
consequence, one can apply either Eq.~(9) 
or (10) to calculate $\Gamma(\Sigma_c^*\to\Lambda_c\pi)$.
From the CLEO measurements \cite{CLEOa}
\be
\Gamma(\Sigma_c^{*++})=\Gamma(\Sigma_c^{*++}\to\Lambda_c^+\pi^+) &=& 
17.9^{+3.8}_{-3.2}\pm 4.0\,{\rm MeV},   \non \\
\Gamma(\Sigma_c^{*0})=\Gamma(\Sigma_c^{*0}\to\Lambda_c^+\pi^-) &=& 
13.0^{+3.7}_{-3.0}\pm 4.0\,{\rm MeV},   
\en
and 
\be
m_{\Sigma_c^{*++}} &=& m_{\Lambda_c}+(234.5\pm 1.1\pm 0.8)\,{\rm MeV}, \non \\
m_{\Sigma_c^{*0}} &=& m_{\Lambda_c}+(232.6\pm 1.0\pm 0.8)\,{\rm MeV},   
\en
we obtain \footnote{The same result for $|g_2|$ is also obtained by D. Pirjol
and T.M. Yan \cite{Yan1}.}
\be
|g_2|=\cases{ 0.61\pm 0.09,  &~~ $\Sigma_c^{*++}\to\Lambda_c^+\pi^+$;   \cr
0.53\pm 0.11, & ~~ $\Sigma_c^{*0}\to\Lambda_c^+\pi^-$.  \cr}
\en
We see that the average experimental value 
\be
|g_2|=0.57\pm 0.10
\en
is in agreement with the quark model prediction $|g_2|
=0.61$, but it deviates $2\sigma$ from the large-$N_c$ argument: 
$|g_2|=0.88\,$ [see Eq.~(5)]. Note that our value of $|g_2|$ is 
slightly different from the result $|g_2|=(0.9\pm 0.2)/\sqrt{3}$ obtained
in \cite{Boyd} due mainly to the presence of the kinematic factor
$(m_{\Lambda_c}/m_{\Sigma_c^*})$ in Eq.~(9).
It is worth mentioning
that $\Xi_c'^*\to\Xi_c\pi^\pm$ decays have also been seen by CLEO with
the results \cite{CLEOb}:
\be
\Gamma(\Xi_c^{'*+})<3.1\,{\rm MeV},~~~m_{\Xi_c^{'*+}}=m_{\Xi_c^0}+(178.2\pm 
0.5\pm 1.0)\,{\rm MeV},   \non \\
\Gamma(\Xi_c^{'*0})<5.5\,{\rm MeV},~~~m_{\Xi_c^{'*0}}=m_{\Xi_c^+}+(174.3\pm 
0.5\pm 1.0)\,{\rm MeV}.
\en
We find $|g_2|<0.64$ from the experimental limit on $\Gamma(\Xi_c^{'*+})$
and $|g_2|<0.84$ from $\Gamma(\Xi_c^{'*0})$. Applying the experimental 
value (14) leads to
\be
\Gamma(\Xi_c^{'*+})=\Gamma(\Xi_c^{'*+}\to\Xi_c^+\pi^0,\Xi_c^0\pi^+) &=& 
(2.44\pm 0.85)\,{\rm MeV},   \non \\
\Gamma(\Xi_c^{'*0})=\Gamma(\Xi_c^{'*0}\to\Xi_c^+\pi^-,\Xi_c^0\pi^0) &=& 
(2.51\pm 0.88)\,{\rm MeV}.
\en
Note that we have neglected the effect of $\Xi_c-\Xi'_c$ mixing in
calculations (for recent considerations, see \cite{Boyd,Ito}).
Therefore, the predicted total decay rate of $\Xi_c^{'*+}$ is very close to 
the current limit $\Gamma(\Xi_c^{'*+})<3.1$ MeV \cite{CLEOb}.

    The decay width of $\Sigma_c$ has not been measured. From Eqs.~(9) and
(14) we obtain
\be
\Gamma(\Sigma_c^0\to\Lambda_c^+\pi^-)=\,(1.94\pm 0.57)\,{\rm MeV}.
\en
It is clear that the strong decay width of $\Sigma_c$ is smaller than that of
$\Sigma_c^*$ by a factor of $\sim 7$, although they will become the same 
in the limit of heavy quark symmetry. This is ascribed to the fact that
the c.m. momentum of the pion is around 88 MeV in the decay 
$\Sigma_c\to\Lambda_c\pi$ while it is two times bigger in 
$\Sigma_c^*\to\Lambda_c\pi$. 

   Thus far we have neglected the electromagnetic contributions to
the total decay width of heavy baryons. To check this, we note that the
radiative decays are described by the amplitudes
\be
A(B_6\to B_{\bar 3}+\gamma) &=& i\eta_1\bar u_{\bar 3}\sigma_{\mu\nu}k^\mu
\ep^\nu u_6,   \non \\
A(B^*_6\to B_{\bar 3}+\gamma) &=& i\eta_2\epsilon_{\mu\nu\alpha\beta}
\bar u_{\bar 3}\gamma^\nu k^\alpha\ep^\beta u^\mu,  \\
A(B^*_6\to B_6+\gamma) &=& i\eta_3\epsilon_{\mu\nu\alpha\beta}
\bar u_6\gamma^\nu k^\alpha\ep^\beta u^\mu,  \non
\en
where $k_\mu$ is the photon 4-momentum, $\ep_\mu$ is the polarization 
4-vector, and the coupling constants $\eta_i$ can
be calculated using the quark model \cite{em}; some of them are
\be
\eta_1(\Sigma_c^+\to\Lambda_c^+)=\,{e\over 6\sqrt{3}}\left({2\over M_u}+{1
\over M_d}\right),&& \eta_2(\Sigma_c^{*+}\to\Lambda_c^+)=\,{e\over 3\sqrt{6}}
\left({2\over M_u}+{1\over M_d}\right),   \non  \\
\eta_3(\Sigma_c^{*++}\to\Sigma_c^{++})=\,{2\sqrt{2}e\over 9}\left({1\over M_u}
-{1\over M_c}\right),&&\eta_3(\Sigma_c^{*0}\to\Sigma_c^0)=\,{2\sqrt{2}e\over 
9}\left(-{1\over 2M_d}-{1\over M_c}\right),   \non \\
\eta_3(\Sigma_c^{*+}\to\Sigma_c^+)=\,{\sqrt{2}e\over 9}\left({1\over M_u}
-{1\over 2M_d}-{2\over M_c}\right),
&&\eta_3(\Xi_c^{'*+}\to\Xi_c^+)=\,{e\over 3\sqrt{6}}
\left({2\over M_u}+{1\over M_s}\right),  \non \\
\eta_3(\Xi_c^{'*0}\to\Xi_c^0)=\,{e\over 
3\sqrt{6}}\left(-{1\over M_d}+{1\over M_s}\right). &&   
\en
Using the formulae \footnote{If the realistic value of $v'$ is taken, Eq.~(20)
will become \cite{em}
\be
\Gamma(B_6(v)\to B_{\bar 3}(v')+\gamma) &=& \eta_1^2\,{k^3\over \pi},  \non \\
\Gamma(B_6^*(v)\to B_{\bar 3}(v')+\gamma) &=& \eta_2^2\,{k^3\over 3\pi}\,{
3m_i^2+m_f^2\over 4m_i^2},   \non \\
\Gamma(B_6^*(v)\to B_6(v')+\gamma) &=& \eta_3^2\,{k^3\over 3\pi}\,{3m_i^2
+m_f^2\over 4m_i^2}.   \non
\en}
\be
\Gamma(B_6(v)\to B_{\bar 3}(v)+\gamma) &=& \eta_1^2\,{k^3\over \pi}\,{m_f\over 
m_i},   \non \\
\Gamma(B_6^*(v)\to B_{\bar 3}(v)+\gamma) &=& \eta_2^2\,{k^3\over 3\pi}\,{m_f
\over m_i},    \\
\Gamma(B_6^*(v)\to B_6(v)+\gamma) &=& \eta_3^2\,{k^3\over 3\pi}\,{m_f\over 
m_i},   \non
\en
the light constituent quark masses
\be
M_u=338\,{\rm MeV},~~~M_d=322\,{\rm MeV},~~~M_s=510\,{\rm MeV},
\en
from the Particle Data Group \cite{PDG}, and $M_c=1.6$ GeV,
we obtain
\be
\Gamma(\Sigma_c^{*++}\to\Sigma_c^{++}\gamma)=1.4\,{\rm keV},&&
\Gamma(\Sigma_c^{*+}\to\Sigma_c^{+}\gamma)=0.002\,{\rm keV},~~
\Gamma(\Sigma_c^{*0}\to\Sigma_c^{0}\gamma)=1.2\,{\rm keV},   \non \\
\Gamma(\Sigma_c^{*+}\to\Lambda_c^{+}\gamma)=147\,{\rm keV},&&
\Gamma(\Sigma_c^{+}\to\Lambda_c^+\gamma)=88\,{\rm keV},   \non \\
\Gamma(\Xi_c^{'*+}\to\Xi_c^+\gamma)=54\,{\rm keV}, &&
\Gamma(\Xi_c^{'*0}\to\Xi_c^0\gamma)=1.1\,{\rm keV}.
\en
Hence, unlike the $D^0$ case where the radiative decay accounts for one third 
of the $D^0$ rate, the branching ratio of the radiative decays of
charmed baryons is at most a few percent.

  Assuming the validity of the quark model relations (1) for different
coupling constants, it follows from (14) that
\be
|g|=\,0.70\pm 0.12\,,~~~~~|g_1|=\,0.93\pm 0.16\,.
\en
Since this value of $g$ is substantially larger than the QCD sum rule 
estimate, it is of great importance to measure $\Gamma(D^*)$ to clarify the 
long-standing issue with $g$.
As mentioned in passing, the parameter $g_1$ cannot be determined 
from the strong decays of heavy baryons. 
It has been advocated in \cite{Savage} that a measurement of
the branching ratio ${\cal B}(\Xi_c^{'*0}\to\Xi_c^0\gamma)$ will
determine $|g_1|$ directly. As pointed out in \cite{SU3}, the radiative decay
$\Xi_c^{'*0}\to\Xi_c^0\gamma$ is forbidden at tree level in SU(3) limit 
[see Eq.~(19)]. In heavy baryon chiral perturbation theory, this radiative 
decay is induced via chiral loops where SU(3) symmetry is broken by the 
light current quark masses. A complete evaluation of leading chiral loops in
\cite{SU3} yields (see Eqs.~(3.31), (3.39), (3.42) and (3.51) in \cite{SU3})
\footnote{ For simplicity we have neglected $\Xi_c-\Xi'_c$ mixing and
$\Delta m$, the mass splitting
between the sextet and antitriplet baryon multiplets, in loop calculations.
The $\Delta m$ correction to chiral effects of $m_q^{1/2}$ type is considered 
in \cite{Savage}, but it is not significant. Note that only one
of the five chiral loop diagrams has been elaborated in \cite{Savage} and
contributions of order $m_q\ln m_q$ are not taken into account there.}
\be
\eta_3(\Xi_c^{'*0}\to\Xi_c^0) &=& -2\sqrt{3\over 2}\,\Bigg[\,{3\over 8}g_1g_2
a_1(-\ep_K+\ep_\eta)+{3\over 2}g_2^2a_2(-\ep_\pi+\ep_K)   \non \\
&&~~+a_2(-\ep_\pi+\ep_K)+{3e\over 128}\,{g_1g_2\over \pi f_\pi^2}(-m_\pi+m_K)
\Bigg],
\en
where 
\be
\ep_{_P}=\,{1\over 32\pi^2}\,{m_P^2\over f_\pi^2}\,\ln{\Lambda_\chi^2\over 
m_P^2},
\en
$\Lambda_\chi$ is a chiral-symmetry breaking scale of order 1 GeV, and 
$a_1,~a_2$
are coupling constants in the chiral Lagrangian for magnetic transitions of
heavy baryons \cite{em,SU3}. It is clear that chiral-loop effects are
nonanalytic in the forms of $m_q^{1/2}$ and $m_q\ln m_q$ ($m_q$ being the 
current quark mass) and that
a measurement of ${\cal B}(\Xi_c^{'*0}\to\Xi_c^0\gamma)$ does
not fix $g_1$ directly unless some model guidance on the parameters $a_1,~a_2$
is given. In the quark model
$a_i$ are simply related to the Dirac magnetic moments of light quarks.
Writing (see Eq.~(4.21) of \cite{SU3})
\be
a_1=-{e\over 3}\beta,~~~~~a_2=\,{e\over 2\sqrt{6}}\beta,
\en
the parameter $\beta$ is related in the nonrelativistic quark model to
the constituent quark mass, $\beta= 1/M_q$ \cite{em}.
A study of radiative decays of the $D$ mesons has confirmed this
expectation \cite{em,rad}. Using Eqs.~(14), (24) and (26), we can 
utilize the measured $|\eta_3(\Xi_c^{'*0}\to\Xi_c^0)|$ to extract $g_1$. 
By identifying (24) with the quark model prediction given by Eq.~(19),
we find two possible solutions: $|g_1|\approx 1.05\,$ and 0.31\,.  We see that
one of the solutions for $g_1$ is in accord with the quark model value 
$g_1=1$. Of course, {\it a priori} there is no reason to
expect that the two different approaches for radiative decays should agree
with each other exactly since SU(3) violation is treated nonperturbatively
in the quark model, while it is calculated in terms of a perturbative
expansion in chiral effective Lagrangian theory. Nevertheless, the
consistency between theory and model is very encouraging.
It follows from (16) and (22) that the branching ratio of $\Xi_c^{'*0}
\to\Xi_c^0\gamma$ is of order $4\times 10^{-4}$. Although it is probably
difficult to measure this level of branching ratio in the near future, 
measurement of $\Gamma(\Xi^{'*0}\to\Xi_c^0\gamma)$ may prove to be the only
way to determine $|g_1|$, as accentuated in \cite{Savage}.

   In summary, the strong coupling constant $g_2$ is extracted from the CLEO
measurement of $\Sigma_c^*\to\Lambda_c\pi$ to be $|g_2|=0.57\pm 0.10$\,.
In heavy baryon chiral perturbation theory, the
radiative decay $\Xi_c^{'*0}\to\Xi_c^0\gamma$ is induced by chiral
loops where SU(3) symmetry is broken by the light current quark masses.
A measurement of $\Gamma(\Xi_c^{'*0}\to\Xi_c^0\gamma)$ will provide two
possible solutions to $g_1$. Assuming this radiative
decay rate is the same as that estimated by
the constituent quark model, we found that one of the solutions $|g_1|\approx
1.05\,$ is precisely what expected from the quark model.
To conclude, the nonrelativistic quark model predictions 
for the strong coupling constants of ground-state heavy hadrons with 
Goldstone bosons: $g=0.75,~g_1=1$ and $g_2=-0.61$, 
are consistent with all existing experiments.

\vskip 2.5 cm
\centerline{\bf ACKNOWLEDGMENTS}
\vskip 0.5cm
   I am grateful to Tung-Mow Yan for discussions and for reading the 
manuscript. This work was supported in part by the National Science Council 
of ROC under Contract No. NSC86-2112-M-001-020.

\vskip 2.5 cm
\renewcommand{\baselinestretch}{1.1}
\newcommand{\bi}{\bibitem}
%

%
\end{document}